\documentclass[referee]{aa} 
\usepackage{graphicx}
\usepackage{natbib}
\bibpunct{(}{)}{,}{}{}{}
\begin{document}
   \title{Self-similar spherical shock solution with sustained
   energy injection}

   \author{V. I. Dokuchaev
          }
    \institute{Institute for Nuclear Research
    of the Russian Academy of Sciences \\
60th October Anniversary prospect 7a, 117312 Moscow, Russia \\
    \email{dokuchaev@inr.npd.ac.ru}
              }

 \date{Received February 19, 2002 / Accepted XXXX XX, 2002}

   \abstract{
We present the generalization of the Sedov-Taylor self-similar strong
spherical shock solution for the case of a central energy source varying
in time, $E=A t^k$, where $A$ and $k$ are constants. The known
Sedov-Taylor solution corresponds to a particular adiabatic case of $k=0$
or \emph{instant shock} with an instant energy source of the shock,
$E=A$. The self-similar hydrodynamic flow in the nonadiabatic $k\neq0$
case exists only under the appropriate local entropy (energy) input which
must be supported by some radiative mechanism from the central engine.
The specific case of $k=1$ corresponds to a permanent energy injection
into the shock, or \emph{injection shock} with a central source of
constant luminosity, $L=A$, $E=A t$.  The generalized self-similar shock
solution may be applied to astrophysical objects in which the duration of
central source activity is longer than the shock expansion time, e.~g.
the early phase of SN explosions, strong wind from stars and young
pulsars, non-steady spherical outflow from black holes and collapsing
dense stellar clusters with numerous neutron star collisions.
   }

   \titlerunning{Self-similar shock with energy injection}

   \maketitle
%

\section{Introduction}

The well known spherical shock solution \citep{sed46,neu47,tay50,sta69}
describes the self-similar expansion of a strong spherical shock
generated by instant deposition of energy $E=const$ by the central source
in a homogeneous gas medium with density $\rho_1=const$. This
\emph{instant shock} solution is commonly used in numerous astrophysics
applications e.g. for the modeling of SN explosions and evolution of
young SN remnants. The corresponding solution for a strong
ultra-relativistic blast wave was obtained by \citet{bla76}. For recent
reviews of astrophysical shock models see \citet{ost88} and \citet{bis95}
and references therein. The basic requirement for the realization of the
instant shock solution is a short-duration injection of energy $E$ into
the shock.  However there exist possible physical situations of permanent
injection of energy into the expanding shock which we call the
\emph{injection shock},  when a central source has some time-varying
luminosity $L=L(t)$. More exactly the instant shock solution is not
applicable when the duration of energy generation by the central source
$t_\mathrm{s}$ is comparable or exceeds the shock expansion time
$t_\mathrm{sh}$, $t_\mathrm{s}\geq t_\mathrm{sh}$. The condition
$t_\mathrm{s}\geq t_\mathrm{sh}$ is typical for the early stage of a SN
explosion, powerful wind from stars and non-steady spherical accretion
onto compact objects. The other example is the injection shock produced in
\emph{hidden neutrino sources} \citep{ber01} by successive multiple
fireballs after numerous neutron star collisions in the dense stellar
cluster in a galactic nucleus prior to its collapse into a massive black
hole. Below we derive the extension of the Sedov-Taylor self-similar
spherical shock solution to the case of varying in time energy injection
by the central source of power form $E=A t^k$, where $A$ and $k$ are
constants. The notations and logistics of ``Fluid Dynamics'' by
\citet[Chapter X, \S106]{ll59} are used in the self-similar solution
derivation.

Let us  consider a strong expanding spherical shock in an ideal gas with
polytropic index (Poisson parameter)
\begin{equation}
\gamma=c_\mathrm{p}/c_\mathrm{v}=const,
\end{equation}
where $c_\mathrm{p}$ and $c_\mathrm{v}$ are the gas heat capacities under
constant pressure and volume respectively.  All values on the forward
side of the shock discontinuity surface (non-perturbed gas side) are
designated by index 1, e.g.  $\rho_1$, $p_1$, and behind the discontinuity
surface (shock cavity side) by index 2, e.g. $\rho_2$, $p_2$.  In a strong
shock the pressure behind the shock $p_2$ far exceeds the pressure in the
non-perturbed gas $p_1$.  The precise definition for a strong shock
(determined from the shock adiabat) is $p_2/p_1\gg(\gamma+1)/(\gamma-1)$
and is similar to condition $u_1>>c_\mathrm{s}$, where $u_1$ is the shock
expansion velocity with respect to the non-moving (non-perturbed) gas and
$c_\mathrm{s}=(\gamma p/\rho)^{1/2}$ is the sound speed in the non-moving
gas. The following relations are valid for a strong shock discontinuity:
\begin{equation}
v_2= \frac{2}{\gamma+1}u_1, \quad \rho_2=\frac{\gamma+1}{\gamma-1}\rho_1,
\quad p_2=\frac{2}{\gamma+1}{\rho_1 u_1^2}. \label{surf}
\end{equation}
These relations are the (outer) boundary conditions for our
 problem.


\section{Self-similar solution ansatz}

The ansatz for self-similar expansion of a strong spherical shock is in
the observation that gas motion after the shock is determined by only two
independent parameters:  the initial gas density $\rho_1=const$ and the
total shock energy $E=const$ because for a strong shock we may put
$p_1=0$. From these two parameters and from two independent variables,
radius $r$ and time $t$ it is possible to construct the only
non-dimensional combination $r\left(\rho_1/E t^2\right)^{1/5}$ \citep[see
e.~g.][]{ll59,sed69,sta69}. As a result the gas motion reveals
self-similar behaviour when different spherical parts of the gas after
the shock evolve under a constant value of this non-dimensional
combination.  The law for the shock radius evolution would be
\begin{equation}
R=R(t)=\beta\left(\frac{E t^2}{\rho_1}\right)^{1/5},
\end{equation}
where the constant $\beta$ itself is determined from the exact solution.
The corresponding velocity of shock expansion with respect to the
non-perturbed gas is $u_1=dR/dt$.  The gas motion behind the shock would
be governed by the non-dimensional \emph{self-similar} variable
\begin{equation}
\xi=\frac{r}{R(t)},
\end{equation}
which is the relative radius with respect to the instant shock radius
$R(t)$.  On the surface of shock discontinuity $\xi=1$.

A helpful hint for finding the generalization of instant shock solution
is in the following observation:  The discussed self-similarity survives
if the shock energy $E$ is the power-law function of time, $E=A t^k$, with
$A=const$ and $k=const$.  This is because the self-similar variable $\xi$
is the power-law function of time.  Putting $E=A t^k$ retains the power
law dependence of the variable $\xi$ on time.  The only complication is in
the changing the power index in the definition of $\xi$ from $2/5$ to
$(2+k)/5$. Now the case $k=0$ corresponds to the known instant shock
solution and, for example the case $k=1$ would correspond to the shock
with a constant luminosity of the shock source, $A=L=const$.  In the
following we find the corresponding self-similar solution of fluid
equations in a general case of arbitrary $k$.  The shock radius in
general case evolves as
\begin{equation}
R= R(t)=\beta\left(\frac{A}{\rho_1}\right)^{1/5}t^{(2+k)/5}. \label{Rt}
\end{equation}
The velocity of shock expansion is
\begin{equation}
u_1= \frac{dR}{dt}=\frac{(2+k)R}{5t}. \label{u1}
\end{equation}
The non-dimensional self-similar variable is defined as
\begin{equation}
\xi=\frac{r}{R(t)}= \frac{1}{\beta}\left(\frac{\rho_1}{A}\right)^{1/5}
\frac{r}{t^{(2+k)/5}}, \label{xi}
\end{equation}


\section{Fluid equations}

We use the set of fluid equations for describing the shocked gas motion,
as follows. The continuity equation in spherical coordinates:
\begin{equation}
\frac{\partial \rho}{\partial t}+ \frac{\partial(\rho v)}{\partial r}+
\frac{2\rho v}{r}=0. \label{contin}
\end{equation}
The momentum conservation (Euler) equation:
\begin{equation}
\frac{\partial v}{\partial t}+ v\frac{\partial v}{\partial r}
-\frac{1}{\rho}\frac{\partial p}{\partial r}=0. \label{mom}
\end{equation}
The entropy equation:
\begin{equation}
\frac{\partial s}{\partial t}+ v\frac{\partial s}{\partial r}= \dot s,
\label{s}
\end{equation}
where $\dot s=\dot s(r,t)$ is the local entropy source (rate of local
entropy production per unit mass).  The self-similar solution with energy
injection (at $k\neq0$) demands the existence of some non-hydrodyna\-mic
``radiative'' mechanism for local energy (and entropy) supply from the
central source into the shocked gas.  In other words it is assumed that
additional energy is radiatively pumped into the shock from the central
source.  For the realization of self-similar motion the entropy source
$\dot s(r,t)$ must be properly tuned.  The required entropy source term
is calculated from the other fluid equations by assuming the self-similar
motion (Eq.~(\ref{s2})). In the particular case of instant shock ($k=0$)
the entropy source term equals zero and the gas motion is adiabatic.

The local energy equation (energy conservation law) is written as
\begin{equation}
\frac{\partial}{\partial t}
\left[\rho(\varepsilon+\frac{v^2}{2})\right]=
-{\rm div}\left[ \rho v(w+\frac{v^2}{2})\right] +\rho T \dot s, \label{E}
\end{equation}
where the gas internal energy density
$\varepsilon=c_\mathrm{v}T=c_\mathrm{s}^2/[\gamma(\gamma-1)]$,
$w=c_\mathrm{p} T=c_\mathrm{s}^2/(\gamma-1)$ is the gas enthalpy density,
$T$ is the gas temperature and $c_\mathrm{s}$ is the sound velocity. The
integral form of the energy conservation equation (\ref{E}) is
\begin{equation}
\frac{\partial }{\partial t}\int
\left[\rho(\varepsilon+\frac{v^2}{2})\right]\,dV= -\oint\left[
\rho\mathbf{v}(w+\frac{v^2}{2})\right]d\mathbf{f} + \int\rho T \dot
s\,dV. \label{Eint}
\end{equation}
From the last equation we find the rate of energy injection into the shock
\begin{equation}
 \dot E=\frac{\partial }{\partial t}\int_0^{R}
 \left[\rho(\varepsilon+\frac{v^2}{2})\right]\, dV=
 \int_0^{R}\rho T \dot s\, dV,
 \label{L}
\end{equation}
where $R=R(t)$ is the outer shock radius. We assume that this energy
injection into the shock is supplied by the central source luminosity
$L=\dot E=kAt^{k-1}$ with the properly tuned local entropy source $\dot
s=\dot s(r,t)$.


\section{Integral of self-similar motion}

The crucial step for finding the strong spherical shock self-similar
solution is the guess of the integral of self-similar motion derived from
the energy balance equation for a chosen sphere expanding in a
self-similar manner by the law $\xi\equiv r/R(t)=const$.  This law of
motion determines according to Eqs.~(\ref{u1}) and (\ref{xi}) the
corresponding radial expansion velocity of this chosen sphere:
\begin{equation}
v_\mathrm{n}=\frac{d}{dt}[\xi R(t)]=u_1 \xi.
\end{equation}
We derive the required integral of self-similar motion in full analogy
with \citet{ll59} procedure for adiabatic ($k=0$) case. The leaking of
energy through the sphere surface $4\pi r^2$ of gas with local velocity
$v$ and density $\rho$ during the small time-interval $\Delta t$ is
\begin{equation}
\rho v(w+\frac{v^2}{2}) 4\pi r^2 \Delta t
\end{equation}
The last equation contains the enthalpy $w=\varepsilon+p/\rho$, where the
gas internal energy density $\varepsilon=c_\mathrm{v}
T=c_\mathrm{s}^2/[\gamma(\gamma-1)]$, because the gas produces some work
under expansion. On the other hand the volume of the sphere increases
during this time-interval $\Delta t$ by the value $4\pi r^2\Delta
tv_\mathrm{n}$.  Inside this volume there is gas with energy
\begin{equation}
  \rho v_\mathrm{n}(\varepsilon+\frac{v^2}{2}) 4\pi r^2\Delta t.
\end{equation}
The central source pumps into this volume by a nonadiabatic process the
additional energy
\begin{equation}
  \rho v_\mathrm{n} T \Delta s 4\pi r^2 \Delta t.
\end{equation}
where $\Delta s=\dot s \Delta t$.  This additional energy  is proportional
to the square of $\Delta t$, and $\Delta t$ is a small time-interval. From
the energy balance equation
\begin{equation}
  \rho v(w+\frac{v^2}{2}) 4\pi r^2 \Delta t =
  [\rho v_\mathrm{n}(\varepsilon+\frac{v^2}{2}) + \rho v_\mathrm{n} T \Delta s ]
  4\pi r^2 \Delta t
\end{equation}
in the thin spherical shell of radius $r<R(t)$, thickness $\Delta r= v_n
\Delta t$ and volume $4\pi r^2\Delta r$ in the limit $\Delta t\rightarrow
0$ we obtain finally the integral of self-similar motion
\begin{equation}
  v(w+\frac{v^2}{2})=v_\mathrm{n}(\varepsilon+\frac{v^2}{2}).
  \label{intmot}
\end{equation}
The left hand side and right hand side of this integral of motion is
simply a different representation of the same energy flux and does not
contain directly the local entropy source $\dot s=\dot s(r,t)$.
Nevertheless this entropy source is taken into account in
Eq.~(\ref{intmot}) indirectly through connections between the
thermodynamic values $w$ and $\varepsilon$ and the entropy relation
$s=\ln(p/\rho^{\gamma})$ for an ideal gas. As a result the required
integral of self-similar motion (\ref{intmot}) has the same functional
form for both the adiabatic $k=0$ an nonadiabatic $k\neq0$ cases.


\section{Solution of self-similar equations}

We will find the self-similar solution for strong spherical shock with
energy injection by using three equations:  the continuity equation
(\ref{contin}), momentum equation (\ref{mom}) and the integral of motion
(\ref{intmot}).  The entropy equation (\ref{s}) will be used for the
determination of the entropy rate source $\dot s$ corresponding to a
suitable injection of energy from the central source.

Let us define the non-dimensional variables $V=V(\xi),\, G=G(\xi),\,
Z=Z(\xi)$ instead of gas velocity $v$, density $\rho$ and sound velocity
$c_\mathrm{s}$:
\begin{equation}
v=\frac{(2+k)r}{5t}V;\quad \rho=\rho_1 G;\quad
c_\mathrm{s}^2=\left[\frac{(2+k)r}{5t}\right]^2Z. \label{func}
\end{equation}
According to Eq.~(\ref{surf}) on the shock discontinuity surface, i.e. at
$\xi=1$, these non-dimensional functions take the values
\begin{equation}
V(1)= \frac{2}{\gamma+1}; \quad G(1)=\frac{\gamma+1}{\gamma-1}; \quad
Z(1)=\frac{2\gamma(\gamma-1)}{(\gamma+1)^2}. \label{surf2}
\end{equation}
The integral of motion from Eq.~(\ref{intmot}) in non-dimensional
variables takes the form
\begin{equation}
Z=\frac{2-(3+\gamma)V+2\gamma{V^2}}{V(V-1)(\gamma V-1)} \label{Z}
\end{equation}
and does not depend on $k$. With this integral of motion the continuity
equation (\ref{contin}) takes the form
\begin{equation}
(1-V)\frac{d\ln G}{d\ln\xi}-\frac{dV}{d\ln\xi} = 3V. \label{cont2}
\end{equation}
Correspondingly the Euler equation (\ref{mom}) can be written in the form
\begin{equation}
\frac{d\ln G}{d\ln\xi}-
\frac{2\!-\!\!2(\gamma\!\!+\!2)V\!\!\!+\!(3\!+\!2\gamma\!+\!\gamma^2)V^2\!\!-
\!2\gamma{V^3}}{(\gamma\!-\!1)(1\!-\!V)V^2(\gamma
V\!-\!1)}\frac{dV}{d\ln\xi}\!=\!
\frac{2\{5\!\!+\!\!V[(2\!+\!k)(V\!\!-\!2)\!+\!\gamma(k\!-\!3)]\}}
{(\gamma\!-\!1)(2\!+\!k)(1\!-V)V}. \label{mom2}
\end{equation}
Eqs.~(\ref{Z}), (\ref{cont2}) and (\ref{mom2}) for functions Z, V, and G
define completely the motion of the shocked gas.  By using these three
equations and the expression $s=c_\mathrm{v}\ln(p/\rho^{\gamma})$ for the
gas entropy we now calculate from Eq.~(\ref{s}) the required entropy
source inside the shocked gas
\begin{equation}
  \dot s(r,t)=\frac{c_\mathrm{v}}{t}\frac{V}{1-V}(\gamma-1)k.
  \label{s2}
\end{equation}
This entropy source is needed for the implementation of self-similar
motion of the shocked gas with energy injection in accordance with the
law of local energy conservation (\ref{E}) or the law of energy input
into the shock (\ref{L}). We will find below from the exact solution that
$2/(\gamma+1)\leq V\leq1/\gamma$. So $ds/dt=0$ in the adiabatic case of
$k=0$ and $ds/dt>0$ in the non-adiabatic case of $k>0$.  See
Fig.~\ref{Fig4} for the radial entropy rate profile for the case of
$k=1$. Meanwhile the entropy source becomes negative, $\dot s<0$, at
$k<0$. It seems that a physically reasonable self-similar solution does
not exist for the $k<0$ case. Resolving the system of Eqs.~(\ref{mom2})
and (\ref{cont2}) with respect to derivatives $dV/d\ln\xi$ and $d\ln
G/d\ln\xi$ we obtain correspondingly:
\begin{equation}
\frac{d\ln V}{d\ln\xi}=\frac{(\gamma V-1) \{10 - 2[2(2+k)-\gamma(k-3)]V +
(3\gamma-1)(2+k)V^2 \}} {(2+k)(1-V)[2-2(\gamma+1)V+\gamma(\gamma+1)V^2] }
\label{dV}
\end{equation}
and
\begin{equation}
\frac{d\ln G}{d\ln\xi}\!=\!
\frac{V\{2(1\!-\!\!V)[1\!+\!(2\gamma\!-\!\!7)V\!+\!4\gamma V^2]\!+\!
k[6\!-\!8(\gamma\!+\!1)V\!\!+\!(7\!\!+\!2\gamma\!+\!5\gamma^2)V^2\!-\!4\gamma
V^3] \}}{(2+k)(1-V)^2[2-2(\gamma+1)V+\gamma(\gamma+1)V^2]}. \label{dlnG}
\end{equation}
Now it is possible to integrate the last two equations separately by using
the boundary conditions from Eq.~(\ref{surf2}).  The integration of
Eq.~(\ref{dV}) is simple but the rather tedious.  We find the solution
for $V(\xi)$ in parametric form:
\begin{eqnarray}
\xi &=& \left\{(\gamma\!+\!1)^2\,
\frac{10\!+\!V[V(2\!+\!k)(3\gamma\!-\!1)\!+
\!2\gamma(k\!-\!3)\!-\!4(2\!+\!k)]}
{2(\gamma-1)(7-\gamma+2k(3+\gamma)]}\right\}^{a_4}
\left[\frac{(\gamma\!+\!1)(\gamma V\!-\!1)}{\gamma\!-\!1}\right]^{a_3}
\times \nonumber \\
&& \left[\frac{(\gamma\!+\!1)V}{2}\right]^{a_2} \left\{
\frac{(\sqrt{D}+A)[\sqrt{D}(\gamma+1)+\gamma^2(k-3)+5\gamma(1+k)-4(2+k)]}
{(\sqrt{D}-A)[\sqrt{D}(\gamma+1)-\gamma^2(k-3)-5\gamma(1+k)+4(2+k)]}
\right\}^{\frac{a_1}{\sqrt{D}}} \label{xigen}
\end{eqnarray}
in the case of $D>0$ and
\begin{eqnarray}
\xi &=& \left\{(\gamma+1)^2\,
\frac{10+V[V(2+k)(3\gamma-1)+2\gamma(k-3)-4(2+k)]}
{2(\gamma-1)[7-\gamma+2k(3+\gamma)]}\right\}^{a_4}\times  \nonumber \\
&&\left[\frac{(\gamma+1)V}{2}\right]^{a_2} \left[\frac{(\gamma+1)(\gamma
V-1)}{\gamma-1}\right]^{a_3} \exp\left(\frac{2a_1}{\sqrt{-D}}\arctan
W\right) \label{xiarc}
\end{eqnarray}
in the case of $D<0$ respectively. In these equations the numerical
constants
\begin{eqnarray} \label{W}
A &=& V(3\gamma -1)(2+k) + k(\gamma-2) - 3\gamma -4; \nonumber \\
D &=& \gamma^2(k-3)^2 + 2(2+k)(9+2k) - 2\gamma(2+k)(9+2k); \nonumber \\
W &=& \frac{(3\gamma-1)(2+k)[2-V(\gamma+1)]\sqrt{-D}}
{a_5\{(2+k)[(3\gamma-1)V-2]+\gamma(k-3)\}-D(\gamma+1)};
\end{eqnarray}
and
\begin{eqnarray}
a_1 &=& \frac{(\gamma\!-\!2)
[3(12\!-\!7\gamma\!+\!13\gamma^2)\!+\!(2\gamma\!+\!1)(3\gamma\!-1\!)k^2]
\!+\!k(33\gamma\!+\!16\gamma^2\!+ \!19\gamma^3\!-\!32)}
{10(2\gamma+1)(3\gamma-1)}; \nonumber \\
a_2 &=& -\frac{(2+k)}{5}; \nonumber \\
a_3 &=& \frac{\gamma-1}{2\gamma+1}; \nonumber \\
a_4 &=&\frac{\gamma[7+k+\gamma(6k-13)]-12-k}
{10(2\gamma+1)(3\gamma-1)}; \nonumber \\
a_5 &=& 4(2+k)-\gamma[5-3\gamma+(5+\gamma)k].
\end{eqnarray}
From Eq.~(\ref{xigen}) and (\ref{xiarc}) it follows that $V(0)=1/\gamma$
is independent of $k$.

Similarly we find the solution for $G(\xi)$ in parametric form by
integrating of Eq.~(\ref{dlnG}).

\noindent Case $D>0$:
\begin{eqnarray}  \label{G}
G\!\!\! &=& \!\!\!\frac{1}{1\!-\!V}\left[\frac{(\gamma\!+\!1)(\gamma
V\!-\!1)}{\gamma-1}\right]^{a_7} \!\left\{(\gamma\!+\!1)^2 \,
\frac{10\!+\!V[V(2\!+\!k)(3\gamma\!-\!1)\!+\!2\gamma(k\!-\!3)\!-\!4(2\!+\!k)]}
{2(\gamma-1)[7-\gamma+2k(3+\gamma)]}\right\}^{a_8} \!\!\times \nonumber \\
&& \left\{
\frac{(\sqrt{D}+A)[\sqrt{D}(\gamma+1)+\gamma^2(k-3)+5\gamma(1+k)-4(2+k)]}
{(\sqrt{D}-A)[\sqrt{D}(\gamma+1)-\gamma^2(k-3)-5\gamma(1+k)+4(2+k)]}\right\}
^{\frac{a_6}{\sqrt{D}}}.
\end{eqnarray}
Case $D<0$:
\begin{eqnarray} \label{Garc}
G &=& \frac{1}{1-V}\left[\frac{(\gamma+1)(\gamma
V-1)}{\gamma-1}\right]^{a_7}\exp\left(\!\frac{2a_6}{\sqrt{-D}}
\arctan W\!\right) \times  \nonumber \\
&& \left\{(\gamma+1)^2 \,
\frac{10+V[V(2+k)(3\gamma-1)+2\gamma(k-3)-4(2+k)]}
{2(\gamma-1)[7-\gamma+2k(3+\gamma)]}\right\}^{a_8},
\end{eqnarray}
where
\begin{eqnarray}
a_6 &=& \frac{3(\gamma+3)\{2+k-\gamma[2+\gamma(k-3)+k]\}}
{(3\gamma-1)(2\gamma+1)}; \nonumber \\
a_7 &=& \frac{3}{2\gamma+1}; \nonumber \\
a_8 &=& \frac{3(\gamma^2+1)}{(2\gamma+1)(3\gamma-1)}.
\end{eqnarray}


\section{The calculation of constant $\beta$ from the energy integral}
\label{beta}

The constant $\beta$ which appears in a self-similar variable $\xi$ can
be found from the shock total energy:
\begin{equation}
E=\int_0^R \left(\rho\frac{v^2}{2}+ \frac{p}{\gamma-1}\right)4\pi r^2dr.
\label{Energy}
\end{equation}
In non-dimensional variables this can be written as
\begin{equation}
\beta^5 \frac{4\pi}{25}(2+k)^2\int_0^1 G\left(\frac{V^2}{2}+
\frac{Z}{\gamma(\gamma-1)}\right)\xi^4d\xi=1. \label{Egen}
\end{equation}
By putting into this integral the calculated functions $ Z,\, V,\, G$
from  (\ref{Z}), (\ref{xigen}) and (\ref{G}) we find after numerical
integration the required $\beta=\beta(\gamma,k)$. This calculation of a
numerical constant $\beta$ completes the finding of generalized
self-similar shock solution. See Table~\ref{Table1} for some examples of
numerical values of $\beta=\beta(\gamma,k)$
\begin{table}
 \centering
 \begin{minipage}{105mm}
  \caption{Numerical values of constant $\beta(\gamma,k)$ for shocks with
  different $\gamma$ and $k=0$ (instant shock with $E=const$), $k=1$ (injection
  shock with $L=const$) and $k=0$ (Shock with $E=At^2$).}
\begin{tabular}{llll}
\hline
 $\gamma=$  & $5/3$ &$7/5$ & $4/3$\\
\hline
 $k=0$  & 1.152  & 1.033 & 0.994  \\
 $k=1$  & 0.929  & 0.826 & 0.793  \\
 $k=2$  &0.368  &  0.288 & 0.271  \\
\hline
\label{Table1}
\end{tabular}
\end{minipage}
\end{table}


\section{Shock with permanent energy injection}

The particular case of the generalized shock solution for astrophysical
application is a permanent energy \emph{injection shock} produced by a
continuous pumping of energy into the shock from the central source of
constant luminosity, $L=const$. It corresponds to the case of $k=1$ and
$E=L t\propto t$ in contrast to the usually applied Sedov-Taylor shock
with $E=const$, which is produced by a instant implosion of energy into
the shock. The possible astrophysical implications of \emph{injection
shock} solutions with $k\neq0$, that is with a central energy source
varying in time, are the early phase of SN explosion, rarefied bubbles in
the interstellar medium after SN explosions, strong wind from stars and
young pulsars, non-steady spherical outflow from accreting black holes
and dense stellar clusters near collapse with frequent neutron star
collisions. Different analytical and numerical approaches were applied by
\citet{fal75,cas75,wea77} to the modeling of permanent energy injection
shocks in the case of stellar winds and on the interstellar medium and
interstellar bubbles.

The external radius of the expanding $k=1$ injection shock evolves with
time according to Eq.~(\ref{Rt}) as
$R(t)=\beta(A/\rho_1)^{1/5}t^{(2+k)/5}\propto t^{3/5}$, where $L=A$ is
the luminosity of the central source and $\rho_1=const$ is the density of
the ambient gas medium.  So the injection shock expands faster than the
instant Sedov-Taylor shock (k=0, $R\propto t^{2/5}$). This is because of
a constant pumping of energy into the shock, $L=const$.  The
corresponding velocity of injection shock expansion is $u_1(t)=
dR/dt=(R/t)(k+2)/5\propto t^{-2/5}$. From the last two equations it is
very clear the physical meaning and the difference between the instant
shock (k=0) and injection shock (k=1) cases respectively is very clear:
\begin{eqnarray}
E &=& [(5/2)^{2}\beta(\gamma,0)^{-5}] R^3 (\rho u^2)=const; \\
L &=& [(5/3)^{3}\beta(\gamma,1)^{-5}]R^2 u (\rho u^2)=const.
\label{Lconst}
\end{eqnarray}
In these expressions $R^3\sim V$ is shock volume and $R^2\sim S$ is the
shock surface.  At $k=0$ we have the expansion law for instant shock
which corresponds to \emph{constant energy} carried by the swept out gas.
The expansion law for a $k=1$ injection shock corresponds to
\emph{constant energy flux} (or constant source luminosity) carried by
the swept out gas.

The discussed strong shock solution is valid only in the region where the
shock expansion velocity $c_\mathrm{s}\ll u(R) \ll c$, where
$c_\mathrm{s}$ is the sound speed in the ambient gas.  The expanding
strong shock becomes weak and disappears when its expansion velocity
drops below the the sound speed $c_\mathrm{s}$.  The maximum radius of
the expanding strong shock $R_\mathrm{sh}$ is obtained from the equality
$u(R_\mathrm{sh})=c_\mathrm{s}$ by using Eqs.~(\ref{Rt}) and ({\ref{u1}):
\begin{equation}
R_\mathrm{sh}=\left[\left(\frac{2+k}{5c_\mathrm{s}}\right)^{2+k}\beta^{5}\,
\frac{A}{\rho_1}\right]^{\frac{1}{3-k}}. \label{Rsh}
\end{equation}
The maximum time of a strong shock expansion is
\begin{equation}
 t_\mathrm{sh}=\left[\left(\frac{2+k}{5c_\mathrm{s}}\beta\right)^{5}
 \frac{A}{\rho_1}\right]^{\frac{1}{3-k}}.
 \label{tsh}
\end{equation}
The minimal radius of Newtonian motion of the shocked gas, which is
called the Sedov length $l_\mathrm{S}$, is defined by equality
$u(l_\mathrm{S})=c$. The  corresponding expression for the Sedov length,
$l_\mathrm{S}=R_\mathrm{sh}(c_\mathrm{s}/c)^{(2+k)/(3-k)}\ll
R_\mathrm{sh}$, is reproduced from Eqs.~(\ref{Rsh}) by substituting $c$
for $c_\mathrm{s}$. So the region of applicability of the strong shock
solution is $l_\mathrm{S}\ll r\ll R_\mathrm{sh}$.


\section{Conclusion}

The derived solution is the generalization of the Sedov-Taylor
self-similar strong spherical shock solution for the case of an energy
injection from the central source of form $E=A t^k$, where $A$ and $k$
are constants. The power-law ansatz $E=A t^k$ only is enough for deriving
the scaling law for shock radius (\ref{Rt}) and shock expansion velocity
(\ref{u1}) accurate to within the numerical constant
$\beta(\gamma,k)\sim1$ without knowing the exact solution. The numerical
value of this constant (see Table~\ref{Table1}) can be calculated from
equation Eq.~(\ref{Egen}) only after the complete solving of the
self-similar problem. The special case of $k=0$ corresponds to the known
Sedov-Taylor solution, while the case $k=1$ corresponds to permanent
energy injection into the shock by a central source of constant
luminosity. The cases with $k<-1$ seem to be nonphysical due to the total
energy divergence at $t\to0$.

The self-similar hydrodynamic flow in the nonadiabatic $k\neq0$ case
exists only under the self-consistency condition~(\ref{s2}) for the local
entropy input. In other words the self-similar behavior of an expanding
shock in the nonadiabatic $k\neq0$ case is realised only under the
appropriate tuning of local entropy (energy) source according to
Eq.~(\ref{s2}). This is the auxiliary physical condition which supposes
some radiative mechanism for sustained energy supply from the central
source into the shocked gas (which depends on the detailed properties of
the cental engine, radiative transfer, gas composition etc). It can be
seen from Figs.~\ref{Fig1} and \ref{Fig4} that the main part of the energy
is injected near the outer boundary of the shock at $\xi\geq0.8$, i.~e. at
the same place where the shocked gas is mainly gathered. The similarities
of the profiles for density and entropy rate are in favor of the principal
realization of the required tuning of the local energy injection
mechanism if the central source radiation absorption would be proportional
to the gas density.

The self-similar shock solution with energy injection may be applied to
the modeling of astrophysical objects in which duration of central source
activity is longer than shock expansion time, such as the early phase of
SN explosion, strong wind from stars and young pulsars, non-steady
spherical outflow from black holes and collapsing dense stellar clusters
with numerous neutron star collisions.

\begin{acknowledgements}
I am grateful to V. S. Berezinsky and B. I. Hnatyk for useful
discussions. This work was supported in part by the INTAS through grant
99-1065.
\end{acknowledgements}

   \begin{figure*}
   \centering
   \includegraphics{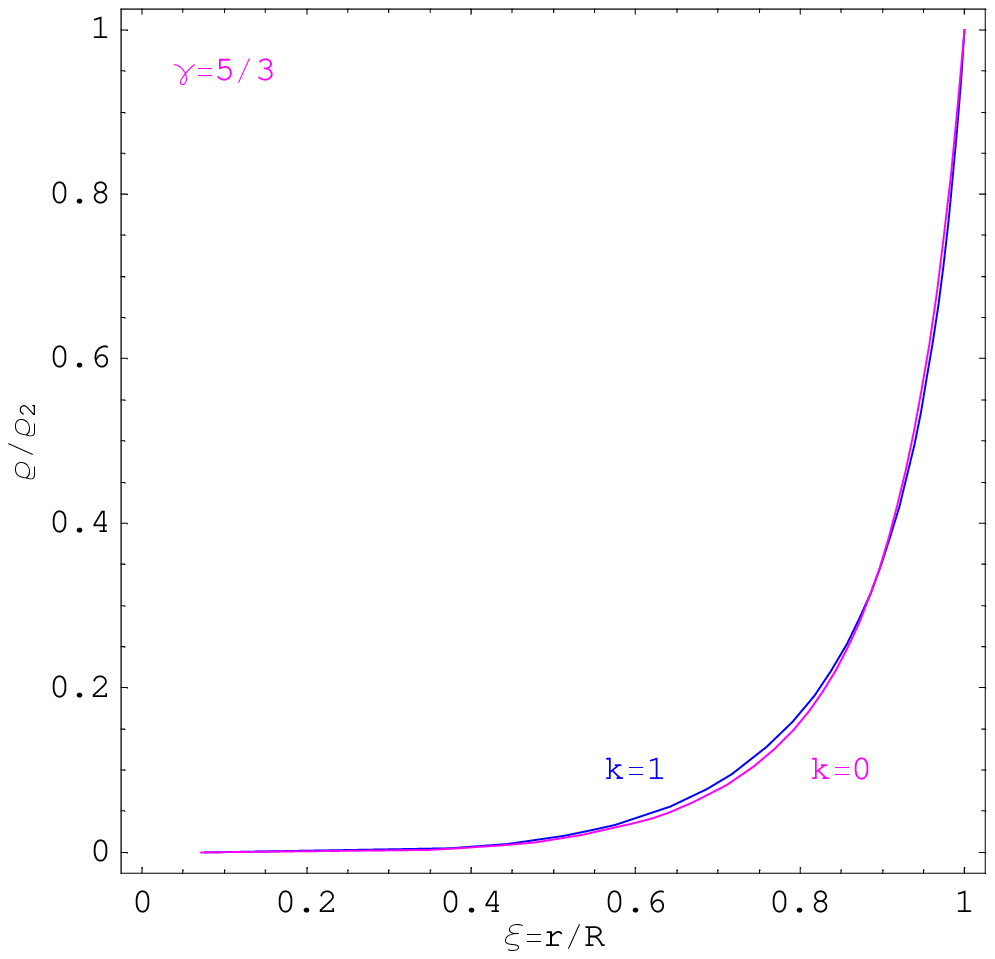}
   \includegraphics{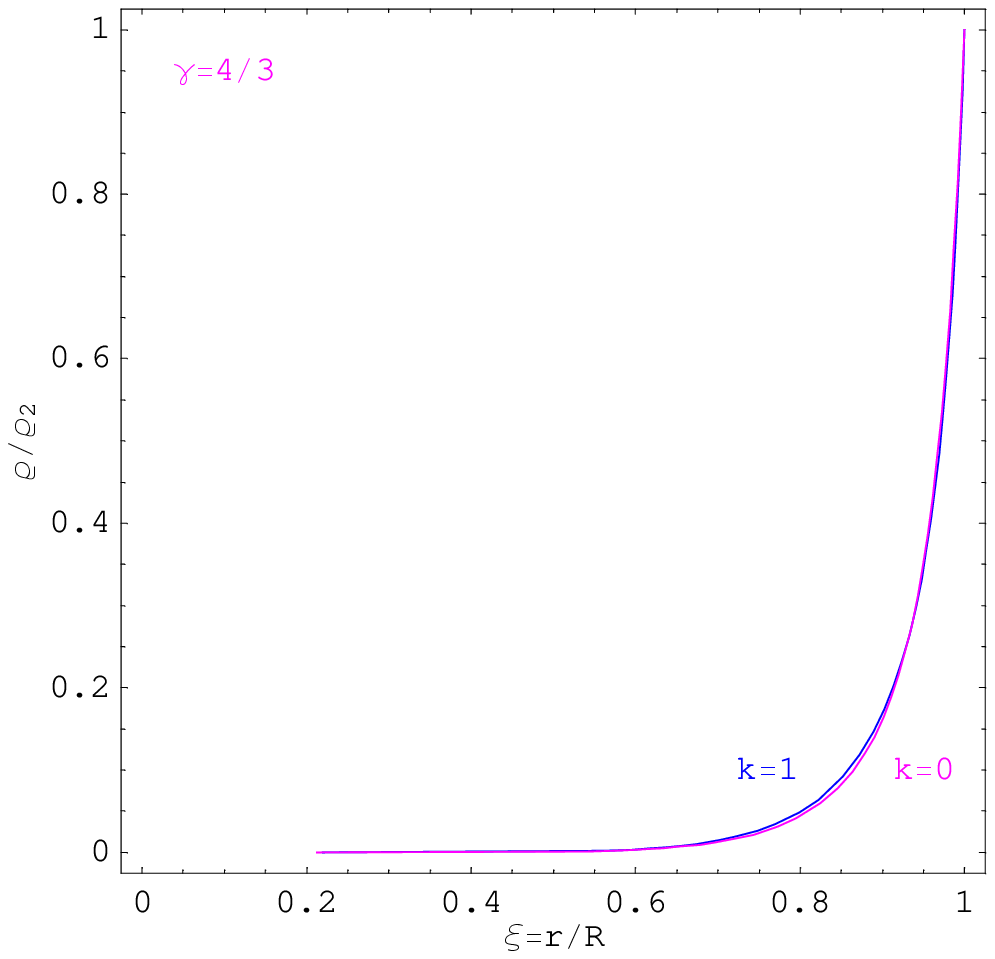}
   \caption{Shock density profile $\rho/\rho_2=G(\xi)(\gamma-1)/(\gamma+1)$
      depending on the relative radius of the shock $\xi=r/R(t)$ for $\gamma=5/3$
      and $4/3$; $k=0$ corresponds to Sedov-Taylor case with the shock energy
      $E=const$; $k=1$ corresponds to the shock with a luminosity of the central
      source $L=const$.}
              \label{Fig1}%
    \end{figure*}
%
   \begin{figure*}
   \centering
   \includegraphics{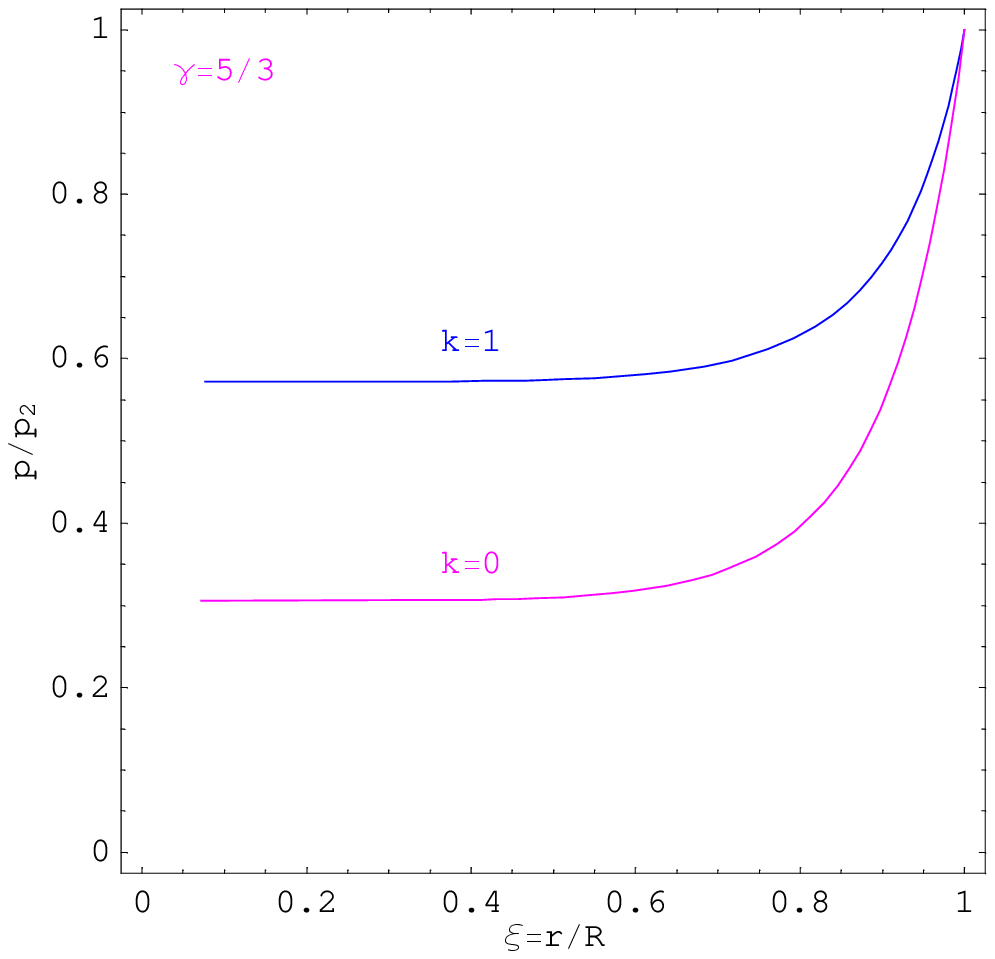}
   \includegraphics{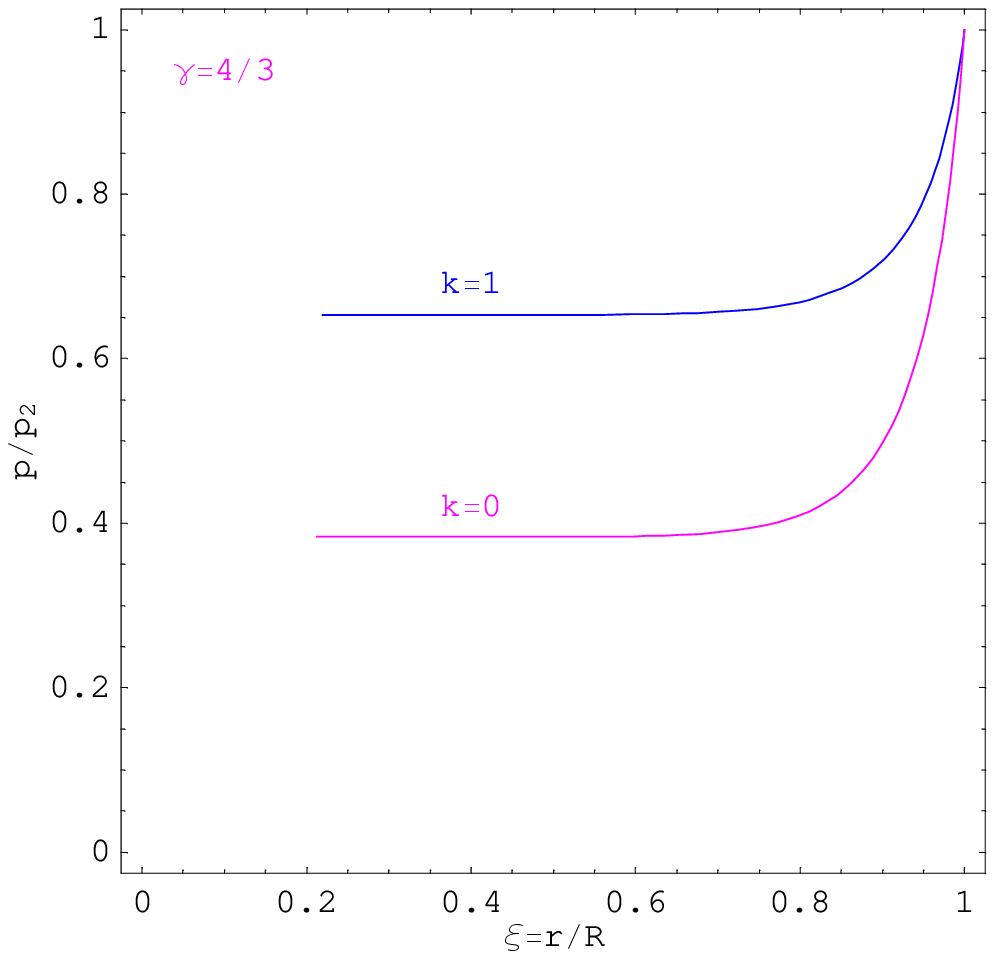}
   \caption{Shock pressure profile $p(r)/p_2=(\frac{\gamma+1}{2\gamma})\xi
G(\xi)Z(\xi)$ depending on the relative radius of the shock $\xi=r/R(t)$
for $\gamma=5/3$ and $4/3$.}
              \label{Fig2}%
    \end{figure*}
%
   \begin{figure*}
   \centering
   \includegraphics{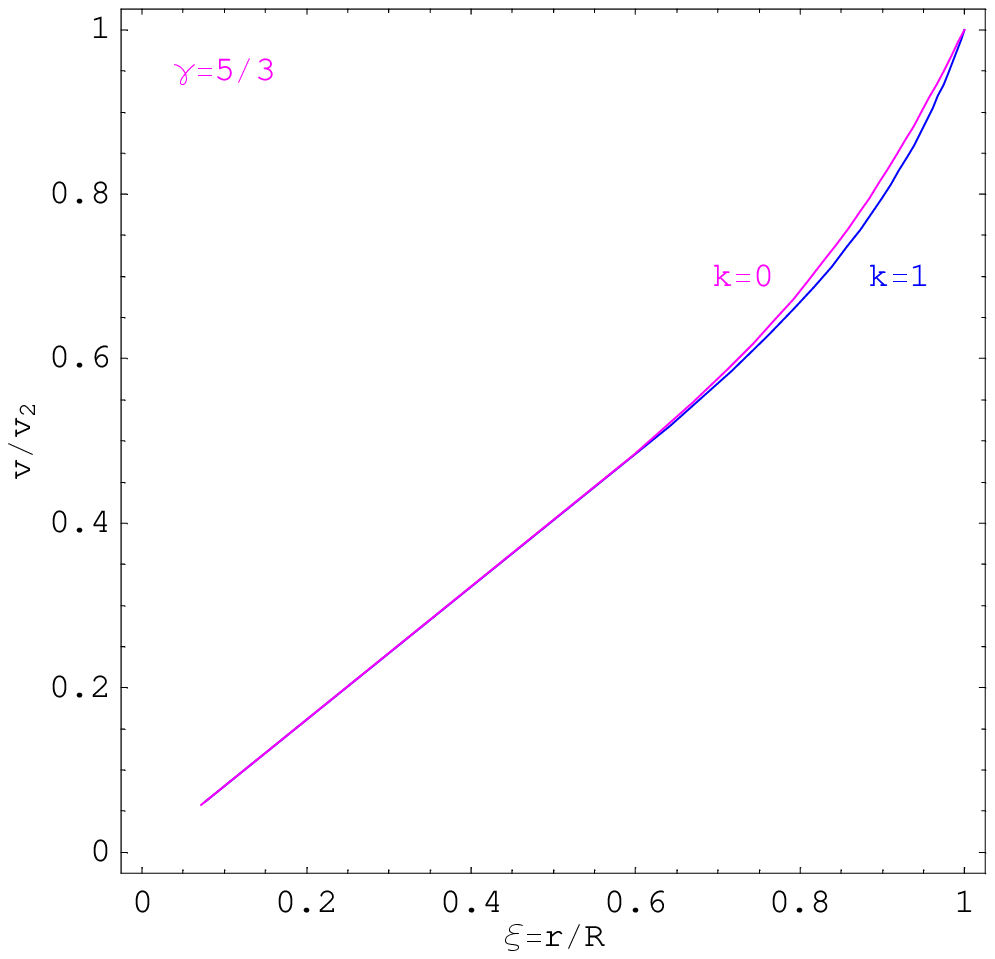}
   \includegraphics{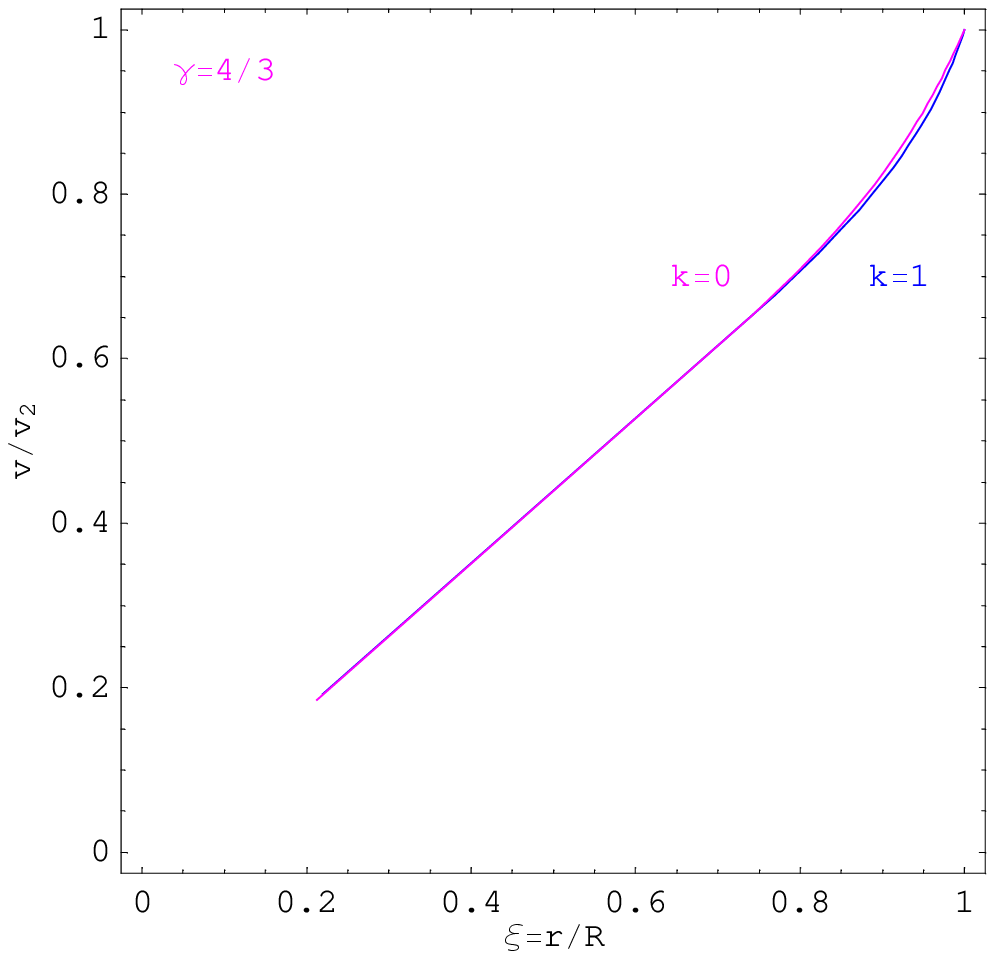}
   \caption{Shock velocity profile $v/v_2=\xi V(\xi)(\gamma+1)/2$
depending on the relative radius of the shock $\xi=r/R(t)$ for
$\gamma=5/3$ and $4/3$.}
              \label{Fig3}%
    \end{figure*}
%
   \begin{figure*}
   \centering
   \includegraphics{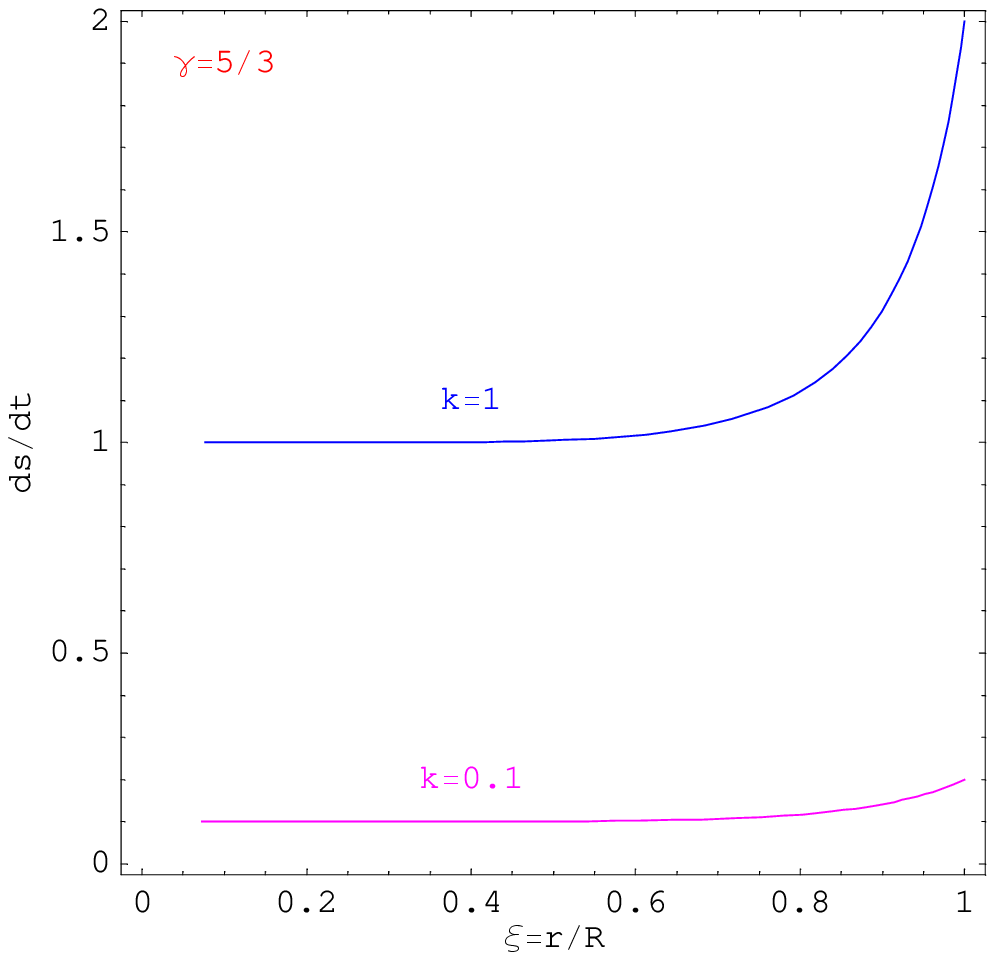}
   \includegraphics{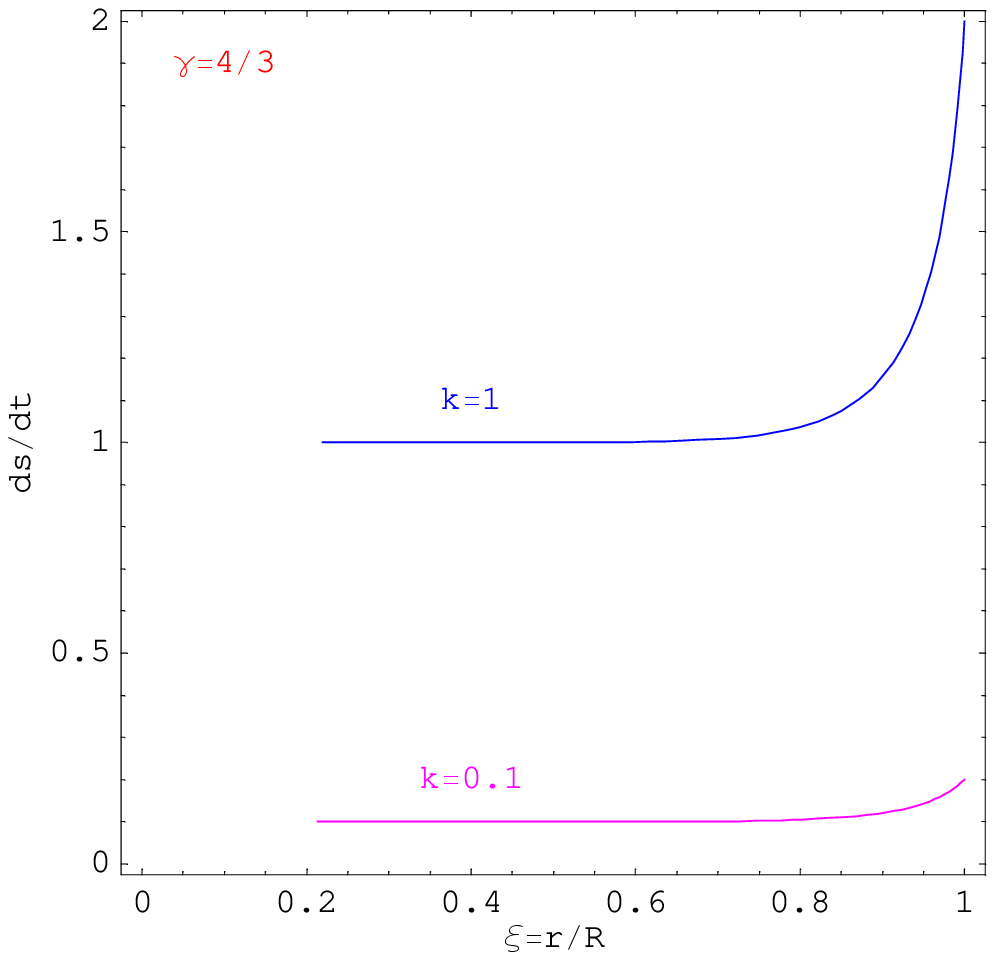}
   \caption{Relative entropy rate $(ds/dt)/(c_\mathrm{v}/t)$ profile, where
$ds/dt$ from Eq.~(\ref{s2}), depending on the relative radius of the
shock $\xi=r/R(t)$ for $\gamma=4/3$ and $5/3$, and $k=0.1$ and $1$.}
              \label{Fig4}%
    \end{figure*}

\end{document}